 \documentclass[final,3p,times]{elsarticle}

\usepackage{amssymb}
\usepackage{amsmath}
\usepackage{hyperref}

\usepackage{graphicx}
\usepackage{dcolumn}
\usepackage{bm}
\usepackage{epstopdf}
\usepackage{bm,color,breakurl}

\journal{Journal of Magnetism and Magnetic Materials}

\begin{document}



\title{Two-spin and multi-spin quantum entanglement\\
in V12 polyoxovanadate molecular nanomagnet}


\author{K. Sza\l{}owski\corref{cor1}}
\ead{karol.szalowski@uni.lodz.pl}
\ead[url]{https://orcid.org/0000-0002-3204-1849}

\address{University of \L\'{o}d\'{z}, Faculty of Physics and Applied Informatics, Department of Solid State Physics,\\ulica Pomorska 149/153, PL90-236 \L\'{o}d\'{z}, Poland}

\cortext[cor1]{Corresponding author}

\begin{abstract}
The paper reports a computational study of the quantum entanglement in V12 cluster molecular magnet. The low-temperature properties of the system are modelled with anisotropic quantum Heisenberg model on a tetramer of spins $S=1/2$ in the external magnetic field. The two-spin entanglement is quantified using the concurrence, whereas the fidelity serves as a measure of four-spin entanglement. The analytic and numerical results are derived and discussed, emphasizing the importance of real-space and spin-space interaction anisotropy and the role of quantum level crossings in the entanglement description in V12. 
\end{abstract}



\date{\today}

\maketitle



\section{Introduction}

The notion of quantum entanglement allows the investigation of genuinely quantum correlations in multi-particle quantum systems and provides a resource for quantum information processing \cite{Horodecki2009}. Molecular magnets focus significant attention as candidate systems for the mentioned purposes \cite{Leuenberger2001,Gaita-Arino2019,Coronado2020}. The description of the molecular magnets uses the spin cluster models with huge diversity of geometries. Within this class of materials, polyoxovanadates offer highly symmetric structures with quantum spins $S=1/2$ forming cage-like clusters \cite{Gatteschi1993}. As an example of polyoxovanadate system, a V12 molecular magnet can be mentioned \cite{Basler2002,Basler2002a}.

The entanglement properties of various cluster magnetic systems were studied in the literature. In the order of increasing cluster size, such systems as: dimers \cite{Tribedi2006,He2006,Candini2010,Abgaryan2011,Yurishchev2011,Reis2012,Das2013,Chakraborty2014,Irons2017,Garlatti2017,Cencarikova2020,Ghannadan2021}, trimers \cite{Wang2001,Bose2005,Tribedi2006,Pal2009,Pal2011,Cima2016}, tetramers \cite{Bose2005,Tribedi2006,Anza2010,Pal2011,Irons2017,Karlova2020}, hexamers \cite{Deb2017} or even larger structures \cite{Wang2002} of various geometry, including spin wheels \cite{Kozlowski2015} of large size and high spin, were investigated, both from the theoretical and experimental point of view. In particular, in the work Ref.~\cite{Bose2005} a discussion of the two-particle entanglement and four-particle entanglement was presented for a tetramer (including V12 case) as a function of the temperature (but without external magnetic field taken into account); also Ref.~\cite{Wang2002} has focused on the tetramer case. It might be mentioned that also the  general thermodynamics of a spin tetramer attracted some attention \cite{Dyszel2021}. On the other side, quantum simulator based on polarized photonic states formally equivalent to a spin tetramer was also discussed \cite{Ma2011}.

In the present work, our aim is to discuss the two-spin and four-spin quantum entanglement in V12 molecular magnet in full parameter space, including the temperature and external magnetic field as an important control parameter. Moreover, the precise inelastic neutron scattering measurements of excitation energies allow the construction of the elaborate magnetic tetramer model with two kinds of anisotropy \cite{Basler2002,Basler2002a}. Therefore, we also point out the important effects of the coupling anisotropy within the V12 tetramer on the entanglement properties, to develop the approach based solely on the isotropic tetramer model in zero field presented in Ref.~\cite{Bose2005}.

In the next part of the paper we briefly characterize the thermodynamics of the model used for V12 system as well as the selected measures of entanglement, whereas in the following section we present our analytic and numerical results. Finally we draw some conclusions from our calculations.

\section{\label{theory}Theoretical model}

\begin{figure}[ht]
\centering
\includegraphics[width=0.7\columnwidth]{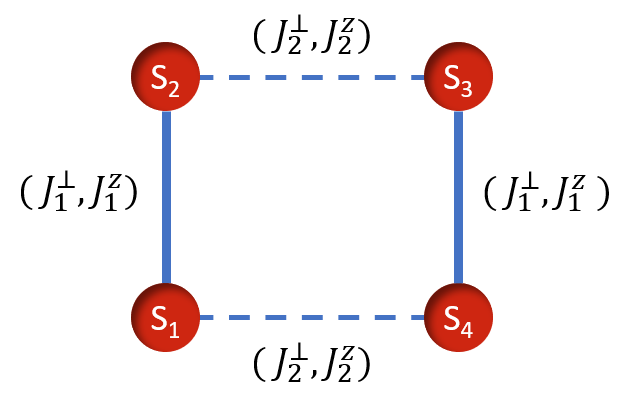}
\caption{\label{sch}The schematic view of spin tetramer used to model V12 cluster molecular magnet. The anisotropic exchange integrals between spin pairs are marked with solid lines (for type I pairs) and with dashed lines (for type II pairs).}
\end{figure}

In the present work, we use the following, most general Hamiltonian of the anisotropic antiferromagnetic Heisenberg-type model for a tetramer composed of spins $S=1/2$ to model the low-temperature properties of V12 \cite{Basler2002,Basler2002a,Szalowski2020}:
\begin{align}\label{hamiltonian}
\mathcal{H}=&-J_1^{\perp}\left(S_1^xS_2^x+S_1^yS_2^y+S_3^xS_4^x+S_3^yS_4^y\right)-J_1^{z}\left(S_1^zS_2^z+S_3^zS_4^z\right)\nonumber\\
&-J_2^{\perp}\left(S_1^xS_4^x+S_1^yS_4^y+S_2^xS_3^x+S_2^yS_3^y\right)-J_2^{z}\left(S_1^zS_4^z+S_2^zS_3^z\right)\nonumber\\&-g\mu_{\rm B}B\left(S_1^z+S_2^z+S_3^z+S_4^z\right).
\end{align}
The above Hamiltonian takes the most general form for which the antiferromagnetic exchange integrals $J_1^{\perp}/k_{\rm B}=$ -18.6 K, $J_1^{z}/k_{\rm B}=$ -19.0 K, $J_2^{\perp}/k_{\rm B}=$ -15.6 K and $J_2^{z}/k_{\rm B}=$ -16.0 K were determined in inelastic neutron scattering experiment reported in Ref.~\cite{Basler2002,Basler2002a}, supplemented with $g=1.97$ \cite{Procissi2004}. In the model, spin-spin couplings are equal along the parallel edges of the rectangle; therefore, two kinds of nearest-neighbour spin pairs are present. In further considerations, we call the pair with interspin coupling $J_1$ a type I pair (spins 1,2 or 3,4), whereas type II pair is coupled with the exchange integral $J_2$ (spins 1,3 or 2,4). For all the spin-spin interactions, the weak spin-space anisotropy of XXZ model type is present, i.e. $J_i^{\perp}\neq J_i^{z}$. The set of four exchange couplings allows the precise reproduction of the diagram of energy levels, clearly showing the presence of the included anisotropies \cite{Basler2002,Basler2002a}. The schematic view of the system in question is shown in Fig.~\ref{sch}.

The fully anisotropic model sketched above allows the precise determination of all the thermodynamic quantities, however, no useful analytic results can be offered and the calculations are only numerically exact. In order to obtain simpler models, the most radical step is taking the equal interspin couplings in the tetramer, leading to $J=J_1^{\perp}=J^{z}_1=J_2^{\perp}=J_2^{z}$. Such a simplest effective theoretical model was sufficient to fit well the temperature dependence of the magnetic susceptibility up to 300 K, as reported in Ref.~\cite{Procissi2004}, with $J/k_{\rm B}=$-17.6 K. Let us mention that the exact forms of all the eigenstates for the tetramer Hamiltonian in this case were given in Ref.~\cite{Wang2002}.

To make better approximation of the physical picture in V12, two different kinds of NN coupling anisotropy can be introduced separately. The first one is taking two different NN couplings, $J_1=J_1^{\perp}=J_1^{z}$ and $J_2=J^{\perp}_2=J^{z}_2$, but still isotropic in spin space. The second one is maintaining the identical interactions for all NN pairs, but introducing an XXZ-type anisotropy in spin space, so that $J^{\perp}=J_1^{\perp}=J_2^{\perp}$ and $J^{z}=J_1^{z}=J_2^{z}$. We will use all the mentioned simplified models for spin tetramer to obtain selected analytic results and to discuss the contribution of individual kinds of anisotropy.

Within the formalism of the canonical ensemble of statistical physics (actually the variant called the field ensemble \cite{Plascak2018}), the density matrix describing the quantum state of the whole system of four spins at the temperature equal to $T$ is of the following form:
\begin{equation}
\rho_{1234}=\frac{1}{\mathcal{Z}}\,e^{-\frac{\mathcal{H}}{k_{\rm B}T}},
\end{equation}
where
\begin{equation}
\mathcal{Z}=\mathrm{Tr}\,e^{-\frac{\mathcal{H}}{k_{\rm B}T}}=\sum_{k=1}^{16}{e^{-\frac{E_k}{k_{\rm B}T}}}
\end{equation}
is the statistical sum taken over all the eigenenergies $E_k$ of the Hamiltonian. 

At the temperature equal to $T=0$ the system takes the ground state of the lowest energy. If the ground state is non-degenerate, it can be described as a pure quantum state which we denote with $\left|\psi_0\right\rangle$ and the corresponding density matrix is $\rho_{1234}=\left|\psi_0\right\rangle\left\langle\psi_0\right|$. The quantity of particular interest for us here is the form of the quantum state being ground state of the system. In general, the states with total spin equal to 0, 1 or 2 can be ground states in various ranges of the magnetic field. The form of the ground states and the expressions for the critical fields is given in \ref{a}; see also the discussion in our work Ref.~\cite{Szalowski2020}). 

If the description of the quantum state of some subsystem within the tetramer is needed, the corresponding density matrix can be obtained by taking the partial trace over the remaining spins not belonging to the specified subsystem. In this spirit, the density matrix for type I NN pair is
\begin{equation}
\rho^{I}_{NN}=\rho_{12}=\rho_{34}=\mathrm{Tr}_{34}\,\rho_{1234}=\mathrm{Tr}_{12}\,\rho_{1234}.
\end{equation}
Analogously, for type II NN pair we have:
\begin{equation}
\rho^{II}_{NN}=\rho_{14}=\rho_{23}=\mathrm{Tr}_{23}\,\rho_{1234}=\mathrm{Tr}_{14}\,\rho_{1234}.
\end{equation}
The pairs of NNN are always equivalent (even in the anisotropic model) and the corresponding density matrix for their state is:
\begin{equation}
\rho_{NNN}=\rho_{13}=\rho_{24}=\mathrm{Tr}_{24}\,\rho_{1234}=\mathrm{Tr}_{13}\,\rho_{1234}.
\end{equation}

In order to quantify the two-spin entanglement within a pair of spins $S=1/2$, we utilize a commonly used entanglement measure called Wootters concurrence \cite{Hill1997,Wootters1998}. It is defined for arbitrary state of a pair of spins $i$ and $j$ as:
\begin{equation}
C_{ij}=\mathrm{max}\left(0,\sqrt{\lambda_1}-\sqrt{\lambda_2}-\sqrt{\lambda_3}-\sqrt{\lambda_4}\right),\label{eq:concurrence}
\end{equation}
where $\lambda_1\geq \lambda_2\geq\lambda_3\geq\lambda_4$ are the eigenvalues of the matrix $\widetilde{\rho_{ij}}\,\rho_{ij}$, with the auxiliary matrix $\widetilde{\rho_{ij}}=\left(\sigma_y\otimes \sigma_y\right)\rho^{*}_{ij}\left(\sigma_y\otimes \sigma_y\right)$. The state $\rho_{ij}$ shows two-particle entanglement if and only if $C_{ij}>0$, with the upper bound for $C_{ij}$ equal to 1. The states with $C_{ij}=0$ are separable. 

In order to quantify the genuine four-particle entanglement, we use the fidelity \cite{Jozsa1994}. The fidelity between the mixed (for example thermal) state $\rho$ and a pure state $\left|\psi\right\rangle$ is defined by
\begin{equation}
F=\left\langle\psi\right| \rho \left|\psi\right\rangle.
\end{equation}
For quantification of the four-spin entanglement, we take the following Greenberger–Horne–Zeilinger (GHZ) state for four spins 1/2 as a reference state showing the genuine four-particle entanglement \cite{Greenberger1989,Greenberger1990,Greenberger2009}:
\begin{equation}
\label{eq:ghz}
\left|\psi_{GHZ}\right\rangle=\frac{1}{\sqrt{2}}\left(\left|\uparrow\downarrow\uparrow\downarrow\right\rangle+\left|\downarrow\uparrow\downarrow\uparrow\right\rangle\right)
\end{equation}
(like it was selected in Ref.~\cite{Bose2005}). It has been verified that this reference state maximizes the fidelity values for the studied thermal states. 

Therefore, the fidelity for four-spin state is defined as
\begin{equation}
F_4=\left\langle\psi_{GHZ}\right| \rho_{1234} \left|\psi_{GHZ}\right\rangle.
\end{equation}
It has been shown that the four-spin state is entangled if $F>1/2$ \cite{Sackett2000}, which sets the criterion for the existence of guaranteed four-particle entanglement.

In the next section we will discuss the analytic and numerical results obtained for quantum entanglement in V12 cluster molecular magnet on the basis of the formalism described above.

\section{\label{num}Results}

In this section we present the results obtained for V12 in both analytic and numerical form, using the formalism described above. All the calculations were performed using the Wolfram Mathematica software \cite{WolframResearchInc.2010}. We present extensively the predictions of the entanglement properties based on the full anisotropic model with Hamiltonian Eq.~\ref{hamiltonian} together with discussion of the analytic results obtained within the simplified models to shed light on the importance of interaction anisotropies. 

\subsection{Two-spin entanglement: Ground-state properties}

\begin{figure}[ht]
\centering
\includegraphics[width=0.7\columnwidth]{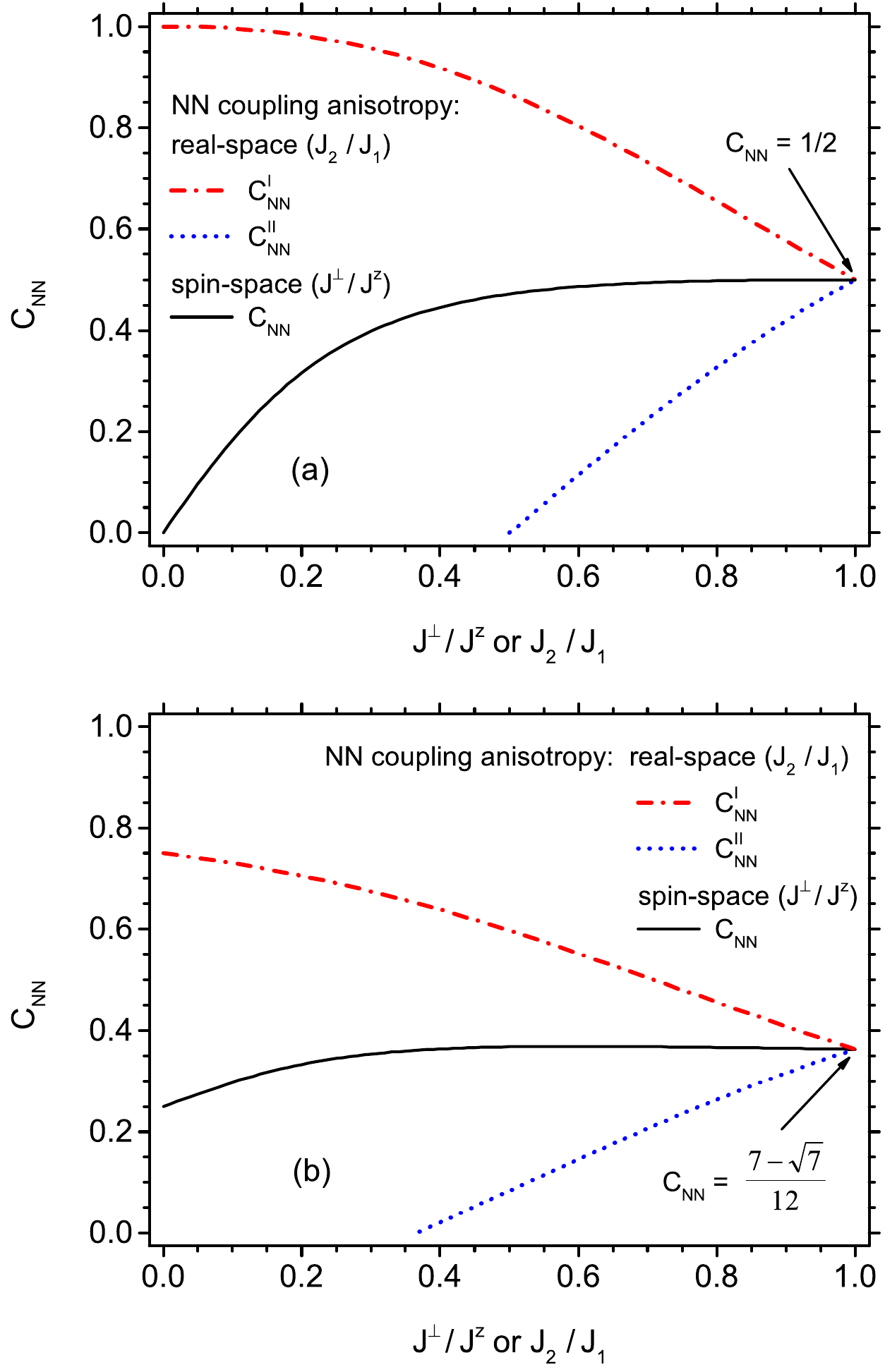}
\caption{\label{anis}The dependence of NN concurrence on the real-space anisotropy or spin-space anisotropy of NN coupling: (a) for the ground state with $S=0$; (b) exactly at the critical field $B_{c,1}$.}
\end{figure}

\begin{figure}[ht]
\centering
\includegraphics[width=0.5\columnwidth]{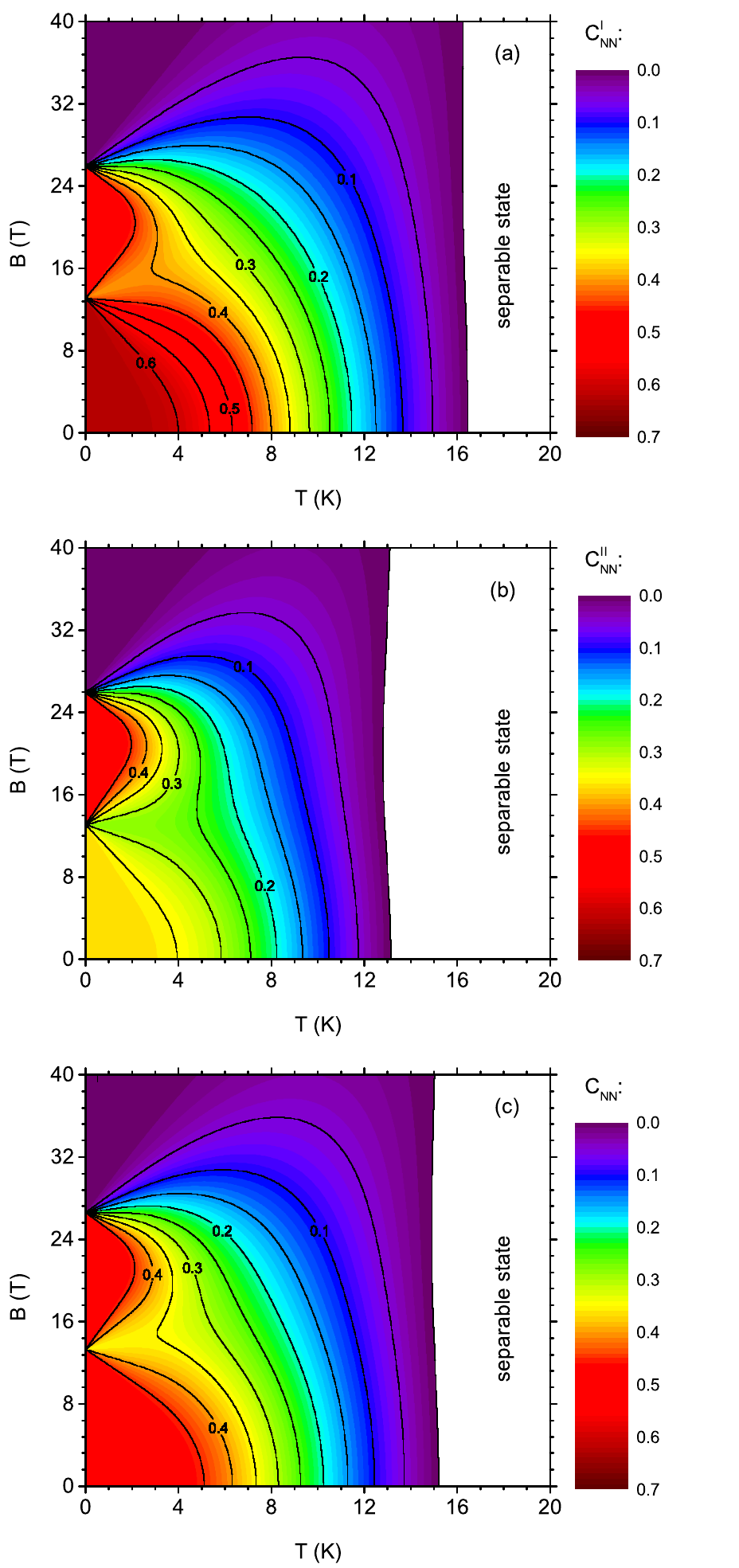}
\caption{\label{cnn}The density plot with contours of the concurrence as a function of the temperature and the external magnetic field: (a) NN spin pairs of type I; (b) NN spin pairs of type II (b) for fully anisotropic tetramer model; (c) NN spin pairs for isotropic tetramer model.}
\end{figure}

Let us commence the discussion from the results concerning the Wootters concurrence, given by Eq.~\ref{eq:concurrence}. The results are discussed separately for ground states with various values of the total spin and for the finite temperature.

The ground state with $S=0$ in V12 is valid for the external magnetic fields below the first critical field $B_{c,1}$ (for detailed discussion see \ref{a} and also Appendix in our Ref.~\cite{Szalowski2020}).

Let us first consider the simplest model of fully isotropic tetramer, for which this ground state takes the form described in \ref{a1}.

In such situation, we arrive at the concurrence value of $C_{NN}=1/2$ for NN spin pairs. Let us mention that exactly such approximation has been discussed in the work Ref.~\cite{Wang2002}.

As mentioned in the section \ref{theory}, a more complicated model for V12 might include two AF couplings different in its magnitude, $J_1$ and $J_2$ (with $|J_1|>|J_2|$), but still isotropic in spin space. Then, two inequivalent NN pairs can be distinguished and the coupling $J_1$ connects type I NN spins, while the coupling $J_2$ is between type II NN spins (see Fig.~\ref{sch}).

For such a model the ground state with spin $S=0$ is discussed in \ref{a2}.

The NN concurrence values based on this state can be expressed analytically with closed formulas, however, we do not present them here in exact form due to their length. As an alternative, we offer the power series expansions with respect to the small parameter $1-\frac{J_2}{J_1}$ around $J_2=J_1$:
\begin{equation}
\label{eq:c1}
C_{NN}^{I}\simeq \frac{1}{2}+\frac{3}{4}\left(1-\frac{J_2}{J_1}\right)+\frac{3}{16}\left(1-\frac{J_2}{J_1}\right)^2-\frac{9}{32}\left(1-\frac{J_2}{J_1}\right)^3-\frac{93}{256}\left(1-\frac{J_2}{J_1}\right)^4.
\end{equation}
and
\begin{equation}
\label{eq:c2}
C_{NN}^{II}\simeq \frac{1}{2}-\frac{3}{4}\left(1-\frac{J_2}{J_1}\right)-\frac{9}{16}\left(1-\frac{J_2}{J_1}\right)^2-\frac{3}{32}\left(1-\frac{J_2}{J_1}\right)^3+\frac{75}{256}\left(1-\frac{J_2}{J_1}\right)^4.
\end{equation}
Therefore, the difference in AF coupling within a tetramer gives rise to unequal values of concurrence for two types of NN pairs, with the larger value of concurrence for stronger coupled spin pair.

In the case of sole spin-space anisotropy of couplings in the studied tetramer, i.e. identical couplings for all NN pairs with $J^{\perp}<J^{z}$, the ground state with spin $S=0$ is given in \ref{a3}.

In this case, the NN concurrence for the ground state discussed here can be expressed by the following useful series expansion with respect to the anisotropy parameter $1-(J^{\perp}/J^{z})$ around $J^{\perp}=J^{z}$:
\begin{equation}
\label{eq:c3}
C_{NN}\simeq \frac{1}{2}-\frac{1}{27}\left(1-\frac{J^{\perp}}{J^{z}}\right)^2-\frac{16}{243}\left(1-\frac{J^{\perp}}{J^{z}}\right)^3-\frac{62}{729}\left(1-\frac{J^{\perp}}{J^{z}}\right)^4.
\end{equation}

Comparison of the effect of both types of interaction anisotropy on NN concurrence below the field $B_{c,1}$ shows that the influence of spin-space anisotropy is generally weaker than the influence of real-space anisotropy. The series expansion Eq.~\ref{eq:c3}  contains no linear term in the parameter $1-(J^{\perp}/J^{z})$ (and the coefficient accompanying a quadratic term is itself rather small), whereas the formulas Eq.~\ref{eq:c1} and Eq.~\ref{eq:c2} include the term linear in $1-(J_2/J_1)$ (with large coefficient). The effect of the anisotropies for the case of NN concurrence is illustrated in Fig.~\ref{anis}(a) (plotted on the basis of full analytic expressions for the plotted quantities). The particularly weak sensitivity of $C_{NN}$ to spin-space anisotropy is well visible. Of course, if the coupling $J^{\perp}\to 0$, the two-spin entanglement vanishes finally, as the ground state becomes $\frac{1}{\sqrt{2}}\left(\left|\uparrow\downarrow\uparrow\downarrow\right\rangle+\left|\downarrow\uparrow\downarrow\uparrow\right\rangle\right)$ (which is the GHZ four-particle state given by Eq.~\ref{eq:ghz}). If a partial trace of this state is taken to calculate the quantum state of a NN pair, it results in separable pair state $\frac{1}{2}\left|\uparrow\downarrow\right\rangle\left\langle\uparrow\downarrow\right|+\frac{1}{2}\left|\downarrow\uparrow\right\rangle\left\langle\downarrow\uparrow\right|$, so that $C_{NN}=0$. The reduction to separable state after taking partial trace over arbitrary subsystem is a general property of the GHZ state \cite{Bengtsson2017}. 

The real-space coupling anisotropy results in appearance of two kinds of NN pairs. It is interesting that for type II NN pair (coupled with $J_2<J_1$), the entanglement vanishes when $J_2/J_1$ reaches a critical value of 1/2. For $J_2/J_1\leq 1/2$ we deal with the case when only the type I NN spin pairs exhibit entanglement. Its measure $C_{NN}^{I}$ tends to 1 when $J_2/J_1$ tends to 0, i.e. in the limit of two uncoupled AF spin dimers. In that limit, the quantum state of each type I NN pair is a Bell state $\frac{1}{\sqrt{2}}\left(\left|\uparrow\downarrow\right\rangle-\left|\downarrow\uparrow\right\rangle\right)$, known to be maximally entangled.

As it follows from the discussion above, the effect of the real-space coupling anisotropy on the NN spin pairs ground-state entanglement is, in general, more pronounced that the influence of spin-space anisotropy.

For the ground state with $S=0$, the concurrence of NNN pairs is $C_{NNN}=0$ for all the models considered, so that the state of NNN pair is separable. 

\begin{figure}[ht]
\centering
\includegraphics[width=0.7\columnwidth]{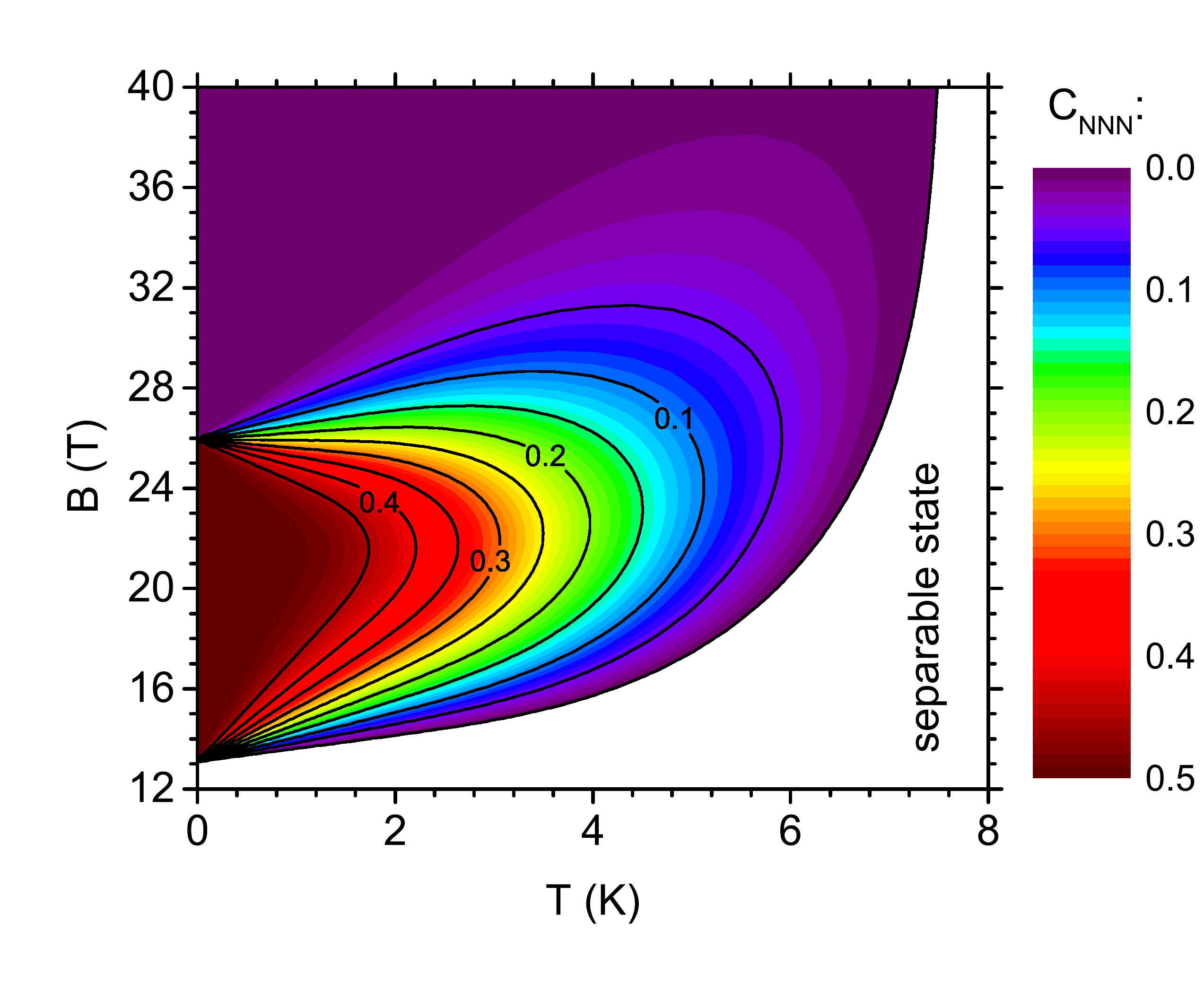}
\caption{\label{cnnn}The density plot with contours of the concurrence as a function of the temperature and the external magnetic field for NNN spin pairs.}
\end{figure}

\begin{figure}[ht]
\centering
\includegraphics[width=0.7\columnwidth]{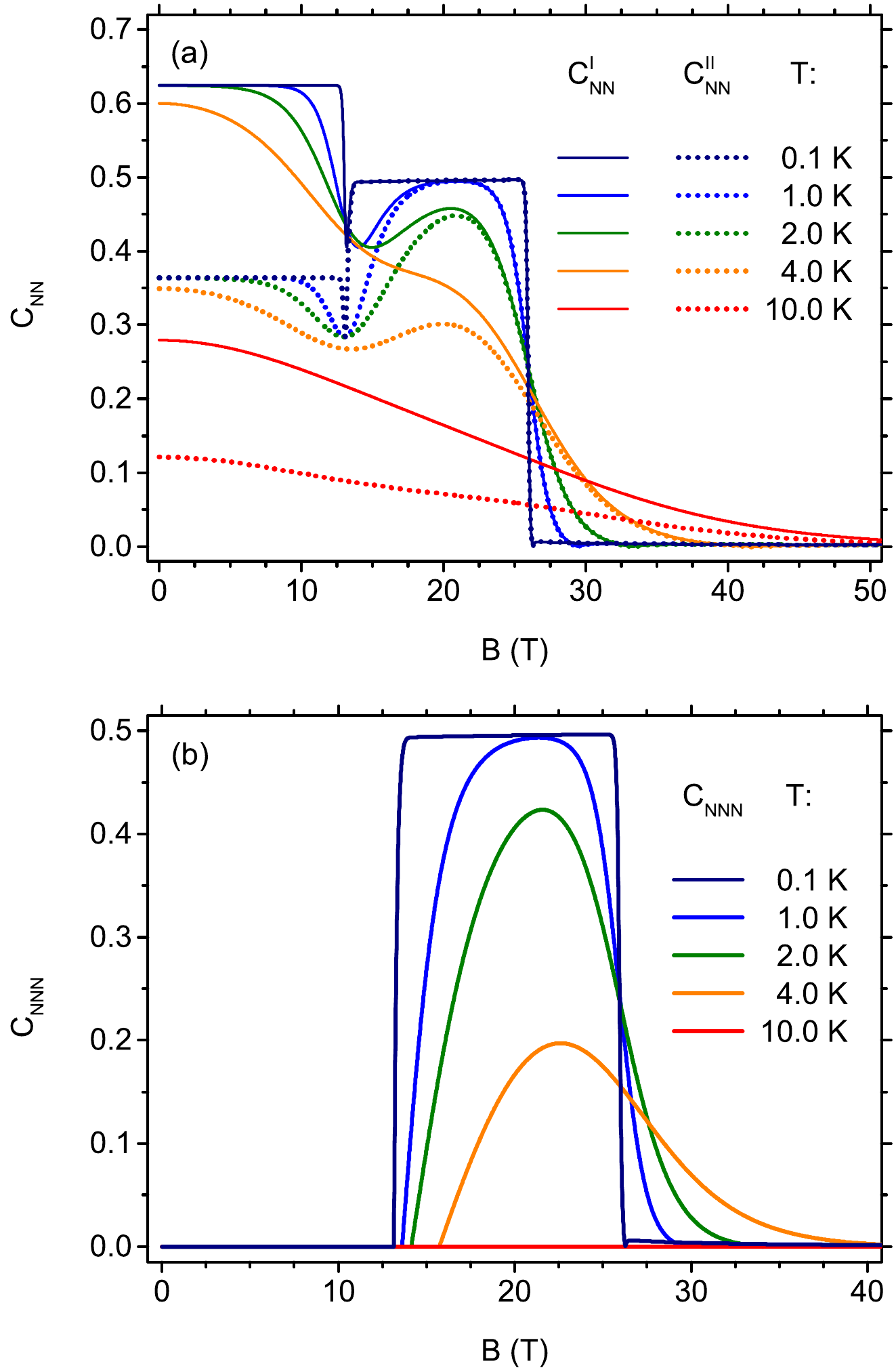}
\caption{\label{cnn2}The dependence of the concurrence on the external magnetic field for selected fixed temperatures, for type I and II NN pairs (a) and NNN pairs (b).}
\end{figure}

\begin{figure}[ht]
\centering
\includegraphics[width=0.7\columnwidth]{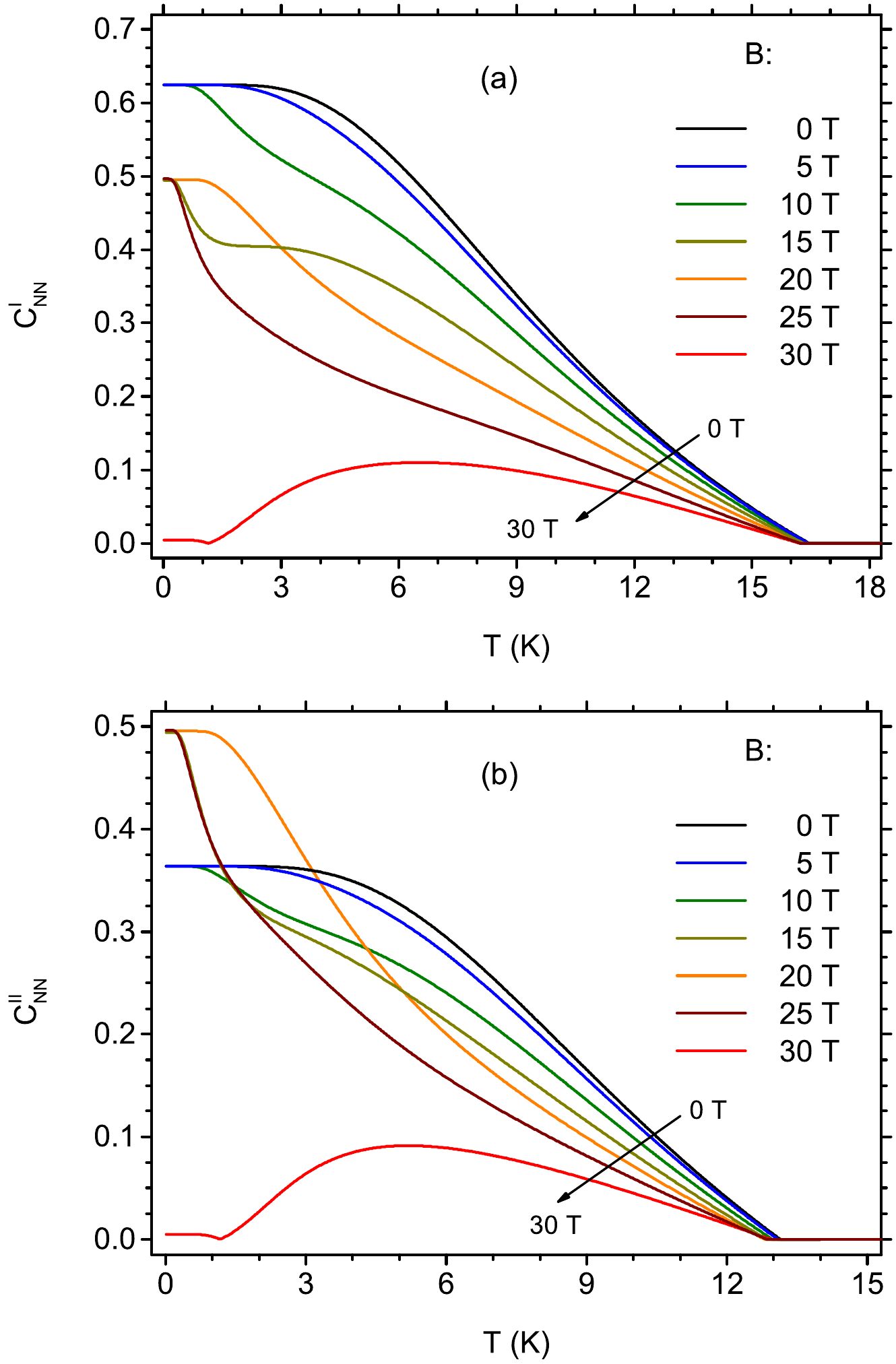}
\caption{\label{cnnt}The dependence of the concurrence on the temperature for selected fixed magnetic fields, for type I NN pairs (a) and type II NN pairs (b).}
\end{figure}

\begin{figure}[ht]
\centering
\includegraphics[width=0.7\columnwidth]{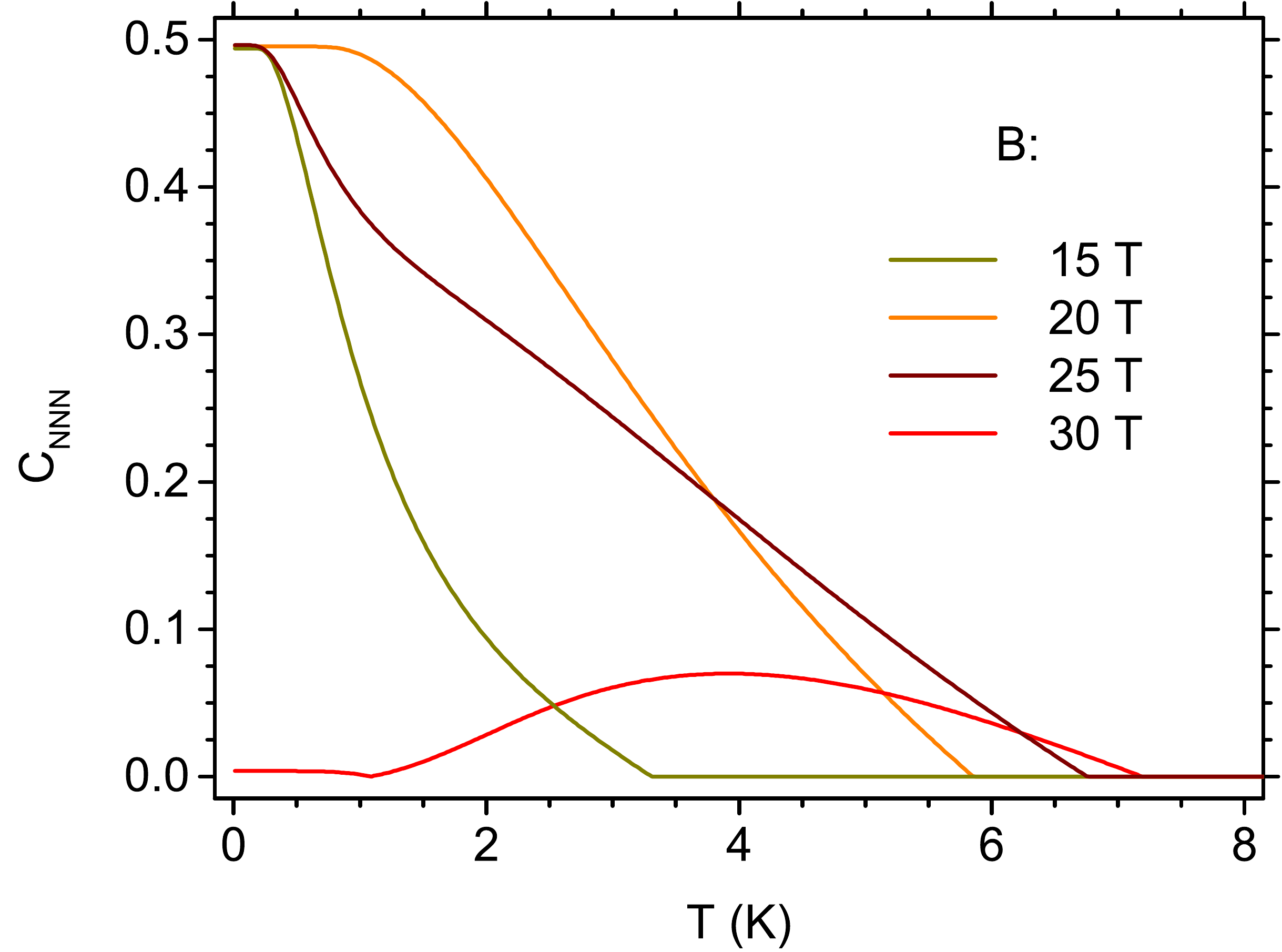}
\caption{\label{cnnnt}The dependence of the concurrence on the temperature for selected fixed magnetic fields for NNN pairs.}
\end{figure}

\begin{figure}[ht]
\centering
\includegraphics[width=0.7\columnwidth]{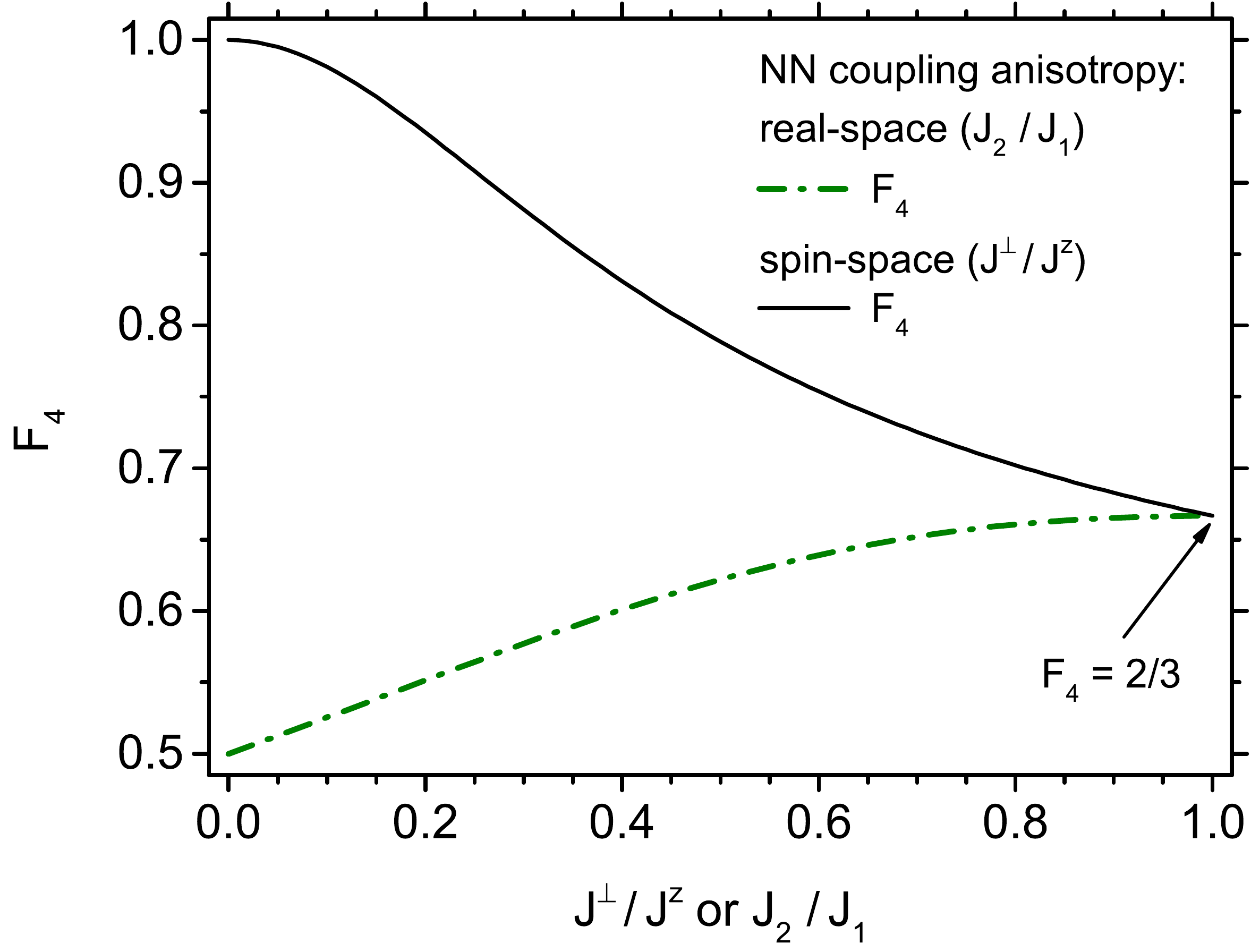}
\caption{\label{f-anis}The dependence of the four-spin fidelity on the real-space anisotropy or spin-space anisotropy of NN coupling for the ground state with $S=0$.}
\end{figure}

\begin{figure}[ht]
\centering
\includegraphics[width=0.7\columnwidth]{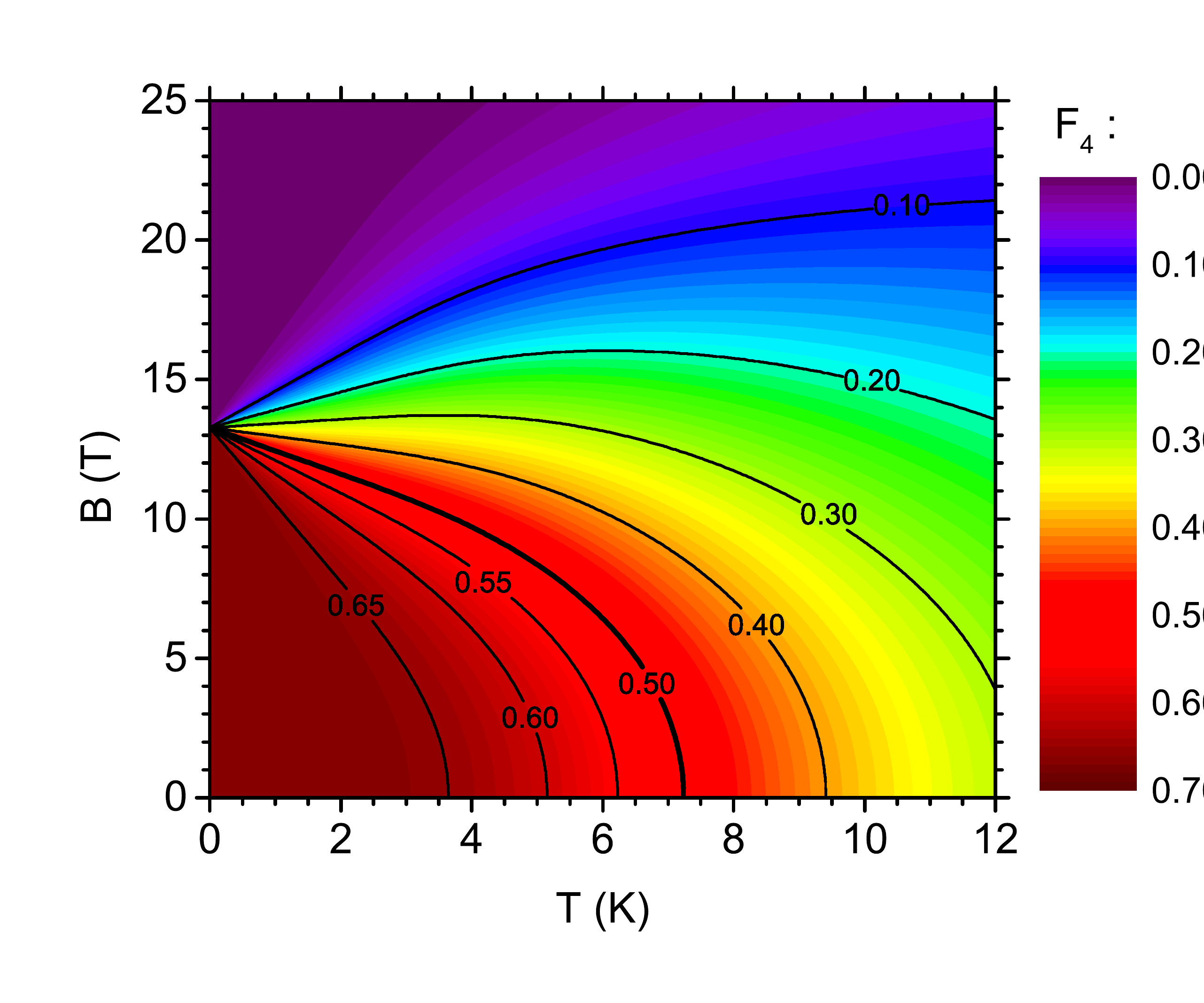}
\caption{\label{fid4}The density plot with contours of the four-spin fidelity as a function of the temperature and the external magnetic field.}
\end{figure}

The ground state with $S=1$ is valid for the external magnetic fields between the first critical field $B_{c,1}$ and the second critical field $B_{c,2}$ (for detailed discussion see \ref{a} and also Appendix in our Ref.~\cite{Szalowski2020}). Its form is given in \ref{a4} (see also Ref.~\cite{Wang2002}). 

For the above state we arrive at the following values of concurrences for all the models considered for the tetramer: $C_{NN}=C_{NNN}=1/2$ .

Exactly for the magnetic field of $B_{c,1}$ the system takes the quantum state of the form $\rho=\frac{1}{2}\left|\psi_{S=0}\right\rangle\left\langle \psi_{S=0}\right|+\frac{1}{2}\left|\psi_{S=1}\right\rangle\left\langle \psi_{S=1}\right|$. The entanglement behaviour at this point deserves a separate discussion. 

If the fully isotropic tetramer model is accepted, the NN concurrence value is $C_{NN}=\frac{7-\sqrt{7}}{12}\simeq 0.362854$. 

Within the model with two unequal couplings isotropic in spin space, the useful power series expansions of the concurrence exactly at the first critical field become:
\begin{equation}
\label{eq:cmix1}
C_{NN}^{I}\simeq \frac{7-\sqrt{7}}{12}+\left(\frac{1}{4}+\frac{\sqrt{7}}{14}\right)\left(1-\frac{J_2}{J_1}\right)+\left(\frac{1}{16}+\frac{55\sqrt{7}}{1568}\right)\left(1-\frac{J_2}{J_1}\right)^2-\left(\frac{3}{32}+\frac{75\sqrt{7}}{21952}\right)\left(1-\frac{J_2}{J_1}\right)^3.
\end{equation}
and
\begin{equation}
\label{eq:cmix2}
C_{NN}^{II}\simeq \frac{7-\sqrt{7}}{12}-\left(\frac{1}{4}+\frac{\sqrt{7}}{14}\right)\left(1-\frac{J_2}{J_1}\right)-\left(\frac{3}{16}+\frac{57\sqrt{7}}{1568}\right)\left(1-\frac{J_2}{J_1}\right)^2-\left(\frac{1}{32}-\frac{47\sqrt{7}}{21952}\right)\left(1-\frac{J_2}{J_1}\right)^3.
\end{equation}

On the other hand, the spin-space anisotropy gives rise to the following value of the NN concurrence at the discussed point:
\begin{equation}
\label{eq:cmix3}
C_{NN}\simeq \frac{7-\sqrt{7}}{12}+\frac{4\sqrt{7}-7}{189}\left(1-\frac{J^{\perp}}{J^{z}}\right)+\frac{4\left(19\sqrt{7}-49\right)}{3969}\left(1-\frac{J^{\perp}}{J^{z}}\right)^2-\frac{2\left(21266-6185\sqrt{7}\right)}{750141}\left(1-\frac{J^{\perp}}{J^{z}}\right)^3.
\end{equation}

The effect of both kinds of coupling anisotropy on the ground-state concurrence at the first critical field can be followed in Fig.~\ref{anis}(b). It is evident that, in similar way as for the case of $S=0$ state, the effect of real-space interaction anisotropy is much stronger than the effect of spin-space anisotropy (even though all the series expansions around the isotropic coupling case include the term linear in anisotropy).

The (ferromagnetically saturated) ground state with $S=2$ in V12 is valid for the external magnetic fields above the second critical field $B_{c,2}$ (for detailed discussion see again \ref{a} and also Appendix in our Ref.~\cite{Szalowski2020}). Its form is shown in \ref{a5} (see also Ref.~\cite{Wang2002}). This state is separable, so that $C_{NN}=0$; moreover the state of NNN is also separable, with $C_{NNN}=0$.

\subsection{Two-spin entanglement: Finite-temperature properties}

In this section, we present the analytic results only for the simplest, fully isotropic model with equal couplings and in the absence of the magnetic field. The temperature dependence of the NN concurrence for this case is given by the formula:
\begin{equation}
\label{eq:ct}
C_{NN}=\frac{1}{2}\frac{1-7 e^{-2\frac{|J|}{k_{\rm B}T}}-10e^{-3\frac{|J|}{k_{\rm B}T}}}{ \left(1+e^{-\frac{|J|}{k_{\rm B}T}}\right) \left[e^{-\frac{|J|}{k_{\rm B}T}} \left(2+e^{-\frac{|J|}{k_{\rm B}T}}\right)+5e^{-2\frac{|J|}{k_{\rm B}T}}\right]}. 
\end{equation}
For $T\to 0$, the value can be approximated with the following expansion:
\begin{equation}
C_{NN}\simeq \frac{1}{2}-\frac{3}{2}e^{-\frac{|J|}{k_{\rm B}T}}-\frac{5}{2}e^{-2\frac{|J|}{k_{\rm B}T}}.
\end{equation}
Both above equations describe monotonic decrease of the NN concurrence with the temperature. Moreover, the critical temperature at which the NN concurrence vanishes can be expressed exactly by:
\begin{equation}
\frac{k_{\rm B}T_{c}}{|J|}=\left[\ln\left( \sqrt[3]{\frac{45+2 \sqrt{249}}{9}}+\frac{\sqrt[3]{135-6 \sqrt{249}}}{3}  \right)\right]^{-1}\simeq 0.863364.
\end{equation}
For the exchange integral $J/k_{\rm B}=-17.6$ K (based on the susceptibility fitting in Ref.~\cite{Procissi2004}), it corresponds to the temperature of 15.2 K.

\subsection{Two-spin entanglement: Numerical results}

Let us discuss now the numerical results for two-spin quantum entanglement calculated numerically on the basis of the fully anisotropic Hamiltonian Eq.~\ref{hamiltonian} for V12, as well as for the case of completely isotropic model for comparison.  

Fig.~\ref{cnn} shows the density plot with contours presenting the concurrence for nearest-neighbour pairs of type I (a) and type II (b) calculated for the fully anisotropic model, supplemented with the NN concurrence (identical for all NN pairs) predicted by the isotropic tetramer model. The concurrence values are plotted in temperature-magnetic field plane. The differences between panels Fig.~\ref{cnn}(a) and Fig.~\ref{cnn}(b) are primarily due to coupling constant difference between type I and the II pairs, in consistency with the discussion contrasting the importance of real-space and spin-space anisotropy. The most pronounced differences can be observed in the range of lower temperatures. For the type I NN pair, the strongest entanglement is noticed for the magnetic fields below the first critical field $B_{c,1}$ (with the values up to 0.6246), i.e. for the field range corresponding to the ground state with $S=0$. The maximum value is larger than 1/2 (and fully consistent with the predictions of Eq.~\ref{eq:c2}). For the type II NN pair, a maximum entanglement with concurrence equal to 0.5 is reached when the magnetic field belongs to the range between the first critical field $B_{c,1}$ and the second critical field $B_{c,2}$, i.e. for the range corresponding to the ground state with spin$S=1$, whereas the concurrence is lower than 1/2 below $B_{c,1}$ (reaching at most the values of 0.3640, consistent with the prediction of Eq.~\ref{eq:c2}). In the range of higher temperatures, two concurrence maxima separated by the critical magnetic field $B_{c,1}$ tend eventually to merge. In the absence of the magnetic field, the temperature at which the entanglement vanishes is equal to 16.5 K for type I pair and to 13.1 K for type II pair. Both values differ significantly from the predictions of the fully isotropic tetramer model (15.2 K). For this case, shown in Fig.~\ref{cnn}(c), two identical maxima of concurrence with the value of 1/2 are visible at low temperatures when the magnetic field increases. The differences between the predictions for type I or type II pairs (Fig.~\ref{cnn}(a) and Fig.~\ref{cnn}(b)) and the   prediction for isotropic  model (Fig.~\ref{cnn}(c)) reflect the important influence of the coupling anisotropy, mainly in real space, on the entanglement in V12 system.

The behaviour of the concurrence for NNN spin pairs as a function of the temperature and external magnetic field can be followed in Fig.~\ref{cnnn}. It is evident there that entanglement of NNN is completely absent at low temperatures for the magnetic fields $B<B_{c,1}$ and it takes maximum values for the fields between the first critical field $B_{c,1}$ and the second critical field $B_{c,2}$, with the largest value of 0.5. If the temperature increases, the maximum becomes smeared, develops some asymmetry and shifts weakly to the larger magnetic fields.

A more detailed investigation of the behaviour of the entanglement requires the analysis of the cross-sections of density plots shown in Fig.~\ref{cnn} and Fig.~\ref{cnnn}. Such a cross-section for various constant values of the temperature for NN pairs is shown in Fig.~\ref{cnn2}, showing the dependence of the concurrence on the magnetic field. The significant difference of the concurrence between type I pair and type II pair (Fig.~\ref{cnn2}(a)) below the magnetic field $B_{c,1}$ is clearly seen. For both pair types, a local minima of concurrence are seen exactly at this field value (with the value in concert with the prediction of Eq.~\ref{eq:cmix1} and Eq.~\ref{eq:cmix2}). For type I (strongly coupled) pair, this minimum shifts to higher fields when the temperature increases, whereas for type II (weakly coupled) pair its position is less sensitive to the temperature. Between $B_{c,1}$ and $B_{c,2}$ both NN concurrences take a common value for the ground state, and if the temperature rises, the difference between both NN pair types becomes increasingly pronounced. In both cases a sort of local maximum is formed, and the increasing temperature causes the entangled range to expand to fields larger than $B_{c,2}$. For high tem,peratures a sort of monotocical decay of concurrence with the field is seen (with still pronounced difference between type I and type II spin pair). The NNN concurrence, shown in (Fig.~\ref{cnn2}(b)), is zero below $B_{c,1}$ at the ground state and if the temperature rises, even the higher field is necessary to trigger the entangled state. At lowest temperatures concurrence for NNN pair vanishes practically above $B_{c,2}$, but thermal fluctuations again tend to diffuse the concurrence maximum, shifting it to higher fields. 

The analysis of the concurrence behaviour as a function of the temperature for the constant magnetic field is possible on the basis of Fig.~\ref{cnnt} for NN pairs and Fig.~\ref{cnnnt} for NNN pairs. For weak fields, the temperature dependence of the concurrence in NN pairs is rather weak at low temperature, similarly to the prediction of Eq.~\ref{eq:ct} valid for isotropic dimer model. Below the magnetic field $B_{c,2}$, the temperature dependence is monotonic, whereas for $B$ exceeding $B_{c,2}$ the behaviour changes qualitatively and a wide temperature-induced maximum is formed. For the range of higher temperatures, the concurrence decreases in linear way with the temperature (and also decreases mononically with the increasing magnetic field). The temperature at which the entanglement vanishes is rather weakly dependent on the magnetic field. The concurrence behaviour for NNN resembles the situation for NN pairs (but the NNN pairs stay unentangled below the magnetic field of $B_{c,1}$.

\subsection{Four-spin entanglement: Ground-state properties}

The quantum state contains surely four-particle entanglement if the fidelity $F_4>1/2$ \cite{Sackett2000}. Such values of the fidelity at the ground state are predicted for the system in question only for the ground state with $S=0$. For the case of fully isotropic spin tetramer, the ground-state fidelity amounts to $F_4=2/3$. If a model with different, but isotropic couplings $J_1$ and $J_2$ is applied, the ground-state fidelity takes the following exact value:
\begin{equation}
\label{eq:f1}
F_4=\frac{1}{3}+\frac{1+\frac{J_2}{J_1}}{6\sqrt{1-\frac{J_2}{J_1}\left(1-\frac{J_2}{J_1}\right)}}.
\end{equation}
A useful expansion of the above expression in power series with respect to the small parameter $1-\frac{J_2}{J_1}$ around $J_2=J_1$ is:
\begin{equation}
\label{eq:f2}
F_4\simeq \frac{2}{3}-\frac{1}{8}\left(1-\frac{J_2}{J_1}\right)^2-\frac{1}{8}\left(1-\frac{J_2}{J_1}\right)^3-\frac{3}{128}\left(1-\frac{J_2}{J_1}\right)^4.
\end{equation}
Therefore, the interaction anisotropy within the tetramer reduces the amount of four-spin entanglement in the ground state with $S=0$. When $J_2$ tends to 0, $F_4$ reaches the limit of 1/2 (as in the limit of two uncoupled dimers there is no four-spin entanglement present in the system).

In the case of only spin-space anisotropy present in the tetramer, the ground-state fidelity is given by the exact expression of the form:
\begin{equation}
\label{eq:fanis}
F_4=\frac{1}{2}\left(1+\frac{1}{\sqrt{1+8\left(\frac{J^{\perp}}{J^{z}}\right)^2}}\right),
\end{equation}
with the following power series:
\begin{equation}
\label{eq:fanisser}
F_4\simeq\frac{2}{3}+\frac{4}{27}\left(1-\frac{J^{\perp}}{J^{z}}\right)+\frac{10}{81}\left(1-\frac{J^{\perp}}{J^{z}}\right)^2+\frac{208}{2187}\left(1-\frac{J^{\perp}}{J^{z}}\right)^3+\frac{1292}{19683}\left(1-\frac{J^{\perp}}{J^{z}}\right)^4.
\end{equation}

The influence of both kinds of anisotropy in model tetramer on the fidelity can be tracked in Fig.~\ref{f-anis}. From the above analysis, it is evident that the influence of the real-space anisotropy on fidelity is weaker, as the expansion with respect to $1-(J_2/J_1)$ (Eq.~\ref{eq:f1}) includes no linear term. On the contrary, the relevant expansion valid for the spin-space anisotropy (Eq.~\ref{eq:fanisser}) involves also a linear term in $1-(J^{\perp}/J^{z})$. The real-space anisotropy weakens the four-spin entanglement, what can be understood by recalling that $J_2\to 0$ means the limit of two non-interacting spin dimers, for which the four-spin entanglement is absent. On the contrary, spin-space anisotropy of XXZ type tends to increase the fidelity, as for the limit of Ising coupling with $J^{\perp}=0$ the exact ground state of the system is $\frac{1}{\sqrt{2}}\left(\left|\uparrow\downarrow\uparrow\downarrow\right\rangle+\left|\downarrow\uparrow\downarrow\uparrow\right\rangle\right)$, which is the reference state  given by Eq.~\ref{eq:ghz} showing maximal entanglement.

\subsubsection{Four-spin entanglement: Finite-temperature properties}

For the case of spin tetramer with equal exchange couplings, the fidelity as a function of the temperature and the external magnetic field is given by the following exact expression:
\begin{equation}
\label{eq:F4-1}
F_4=\frac{2}{3}\frac{\left(1+\frac{1}{2}e^{-3/t}\right) e^{-2h/t}}{e^{-3/t}\left(e^{-4h/t}+1\right)+e^{-1/t}\left(e^{-1/t}+1\right)^2 \left(e^{-3h/t}+e^{-h/t}\right)+\left(e^{-3/t}+3 e^{-2/t}+e^{-1/t}+1\right)e^{-2h/t}},
\end{equation}
where $t=k_{\rm B}T/|J|$ is the dimensionless temperature and $h=g\mu_{\rm B}B/|J|$ is the dimensionless magnetic field. In the absence of the magnetic field, the variability of the fidelity with the temperature reduces to the form (Ref.~\cite{Bose2005}):
\begin{equation}
F_4= \frac{2}{3}\frac{1 +\frac{1}{2}e^{-3/t}}{1+3 e^{-1/t}+7 e^{-2/t}+5e^{-3/t}}.
\end{equation}
The useful low-temperature expansion of the above formula is:
\begin{equation}
F_4\simeq \frac{2}{3}-2e^{-\frac{|J|}{k_{\rm B}T}}+\frac{4}{3}e^{-2\frac{|J|}{k_{\rm B}T}}.
\end{equation}

The borderline $F_4=1/2$ can also be determined exactly and it is given by the following equation:
\begin{equation}
\frac{g\mu_{\rm B}B}{|J|}=\frac{\ln \left(\frac{(x+1)^2+\sqrt{x^4+\frac{16 x^3}{3}+2 x^2-8 x+\frac{23}{3}}}{4} +\frac{\sqrt{3 x^4+14 x^3+12 x^2-6 x-11-(x+1)^2 \sqrt{9 x^4+48 x^3+18 x^2-72 x+69}}}{2 \sqrt{6}}\right)}{\ln x},
\end{equation}

where $x=e^{|J|/\left(k_{\rm B}T\right)}$.

For the range of low temperatures, the expansion of the above formula predicts the linear decrease of the critical magnetic field at which $F_4=1/2$ with the temperature, in the form of:
\begin{equation}
\frac{g_{e}\mu_{\rm B}B}{|J|}\simeq 1-\frac{k_{\rm B}T}{|J|}\ln 3.
\end{equation}
On the other hand, the value of the critical temperature at which $F_4=1/2$ at $B=0$ is expressed by the equation
\begin{equation}
\label{eq:fid4-2}
\frac{k_{\rm B}T}{|J|}=\left[\ln\left(\sqrt[3]{65+\sqrt{129}}+\frac{1}{3} \sqrt[3]{1755-27 \sqrt{129}}+3\right)\right]^{-1}\simeq 0.4168.
\end{equation}
For the exchange integral $J/k_{\rm B}=-17.6$ K (after Ref.~\cite{Procissi2004}), it corresponds to the temperature of 7.34 K.

\subsection{Four-spin entanglement: Numerical results for fully anisotropic model}

The density plot of the four-spin fidelity calculated for the model with all interaction anisotropies included, as a function of the temperature and the external magnetic field, is shown in Fig.~\ref{fid4}. A bold solid line marks the borderline $F_4=1/2$. It is evident that the genuine four-spin entanglement is guaranteed only in the low-temperature range (below 7.24 K; slightly lower value than the prediction of 7.34 K based on the fully isotropic tetramer model of Eq.~\ref{eq:fid4-2}) and below the field of 13.1 T. The maximum value of fidelity amounts to 0.6669, therefore, it is slightly increased with respect to the value of 2/3 predicted for isotropic tetramer model. It can be observed that the real-space and spin-space anisotropies act in opposite directions on the ground state fidelity value (as shown in Fig.~\ref{f-anis}); therefore, the value predicted in the full model is very close to 2/3.

\section{\label{final}Summary and conclusion}

In the paper the entanglement properties of V12 cluster molecular magnet were studied from the computational viewpoint. The two-spin entanglement quantified by concurrence and four-spin entanglement characterized by fidelity were calculated as a function of the temperature and the external magnetic field. 

The discussed spin models for low-temperature magnetism of V12 were based on a quadratic tetramer structure with NN interactions of all-antiferromagnetic type. A spin tetramer can include two basic kinds of NN interaction anisotropy: the real-space anisotropy (leading to formation of two inequivalent types of NN spin pairs) and spin-space anisotropy of XXZ type. The effect of both sorts of anisotropy on the entanglement has been investigated. The most precise model involved both anisotropies and used exchange integrals taken from the experiment \cite{Basler2002}.

For NN concurrence (two-spin entanglement measure), the spin-space anisotropy was found to reduce weakly (quadratically) the entanglement for the ground state with $S=0$. The real-space anisotropy has been predicted to increase the concurrence more significantly (linearly) for NN pairs with stronger interaction and to reduce the concurrence (linearly) for NN pairs with weaker interaction. Therefore, for weak anisotropies, the dominant contribution comes form the real-space anisotropy. The NN concurrence for the ground state with $S=1$ (only at the magnetic field between the first and the second critical field) is insensitive to the coupling anisotropies, so is the form of the ground state with spin $S=1$. 

For four-spin entanglement, the spin-space anisotropy has been found to increase its measure (fidelity) linearly, whereas the real-space anisotropy acts in the opposite direction, reducing the fidelity quadratically (for weak anisotropy).

The present work suggests that the range of temperatures for V12 within which it exhibits significant two-spin entanglement (concurrence above, say, 1/2) is below 6 K for the absence of the magnetic field for type I NN spin pairs, whereas the quantum state is not separable below, approximately, 16 K. The application of the field not exceeding the first critical field $B_{c,1}$ of about 13.3 T reduces the concurrence down to the value of about 0.36. Then, another maximum of the NN entanglement is achieved in the presence of the magnetic field, in between the critical fields $B_{c,1}$ (13.3 T) and $B_{c,2}$ (26.1 T), and is associated with lower concurrence values, below 1/2, due to real-space coupling anisotropy. For higher fields, the ground state is essentially separable and only the thermal admixture of other states leads to some residual entanglement at finite temperatures. The entanglement within 2NN pairs (along the diagonal of the tetramer) is predicted only between the fields $B_{c,1}$ and $B_{c,2}$, with the maximum value of 1/2,  also for the temperatures of the order of few K.  

The genuine four-spin entanglement, quantified by the fidelity (with GHZ state as a reference quantum state) is predicted to be present below approximately 7 K in the absence of the magnetic field. The presence of the field reduces the temperature range for this sort of entanglement, which may be only present below the first critical field of approximately 13 T. 

The properties discussed above may suggest that the V12 molecular magnet might carry both two-spin and four-spin entanglement in the cryogenic range of temperatures (of the order of few K).

The interaction anisotropies in spin tetramer were found to exert somehow opposite effect on the two-spin and four-spin entanglement measures. The strong XXZ anisotropy favours the GHZ four-spin ground state with high genuine four-spin entanglement, whereas it causes the two-spin entanglement to fade out. On the other side, difference in coupling energy for two types of NN pairs tends to reduce the four-spin entanglement, whereas it boosts up the two-spin entanglement (leading finally to the limit of uncoupled dimers, each dimer in maximally entangled Bell state). As a consequence, the selection of the most favourable interactions strongly depends on the desired presence of either type of entanglement. In V12 both two-spin and four-spin entanglement is predicted to be present at cryogenic temperatures and the wide range of magnetic fields. On the other hand, the results regarding the influence of the interaction anisotropies on various aspects of entanglement developed in the paper might be a guide for tuning the possible systems to achieve the desired properties (an example of tuning might be provided by chemical tuning \cite{Siloi2012}). 

It should be mentioned here that the in-deep assessment of the usefulness of V12 for the purpose of quantum information processing would demand a study of such factors as various mechanisms of spin relaxation, as the spins in V12 clusters interacts with their environment. Therefore, intercluster interactions as well as interactions of purely magnetic subsystem with the lattice environment, resulting in decoherence, should be investigated \cite{Ghirri2015}. These topics are outside the scope of the present study. However, it can be commented that polyoxometalate systems, including polyoxovanadates, were considered as favourable candidate systems for quantum information processing \cite{Lehmann2007,Clemente-Juan2012,Ghirri2015,Baldovi2017}.

Finally, let us emphasize that the quantum entanglement properties have recently become  accessible experimentally via elaborate methods based on neutron scattering \cite{Garlatti2017,Scheie2021,Laurell2021}, applied to spin chain systems. This factor gives some hope for similar studies dedicated to cluster systems in the future. Another known system which can be modelled using the spin tetramer Hamiltonian might be, for example, synthetic libethenite Cu$_2$PO$_4$OH \cite{Belik2007} with stronger AF couplings ($J/k_{\rm B}$ of the order of 138 K in fully isotropic tetramer model).




\begin{thebibliography}{10}
\expandafter\ifx\csname url\endcsname\relax
  \def\url#1{\texttt{#1}}\fi
\expandafter\ifx\csname urlprefix\endcsname\relax\def\urlprefix{URL }\fi
\expandafter\ifx\csname href\endcsname\relax
  \def\href#1#2{#2} \def\path#1{#1}\fi

\bibitem{Horodecki2009}
R.~Horodecki, P.~Horodecki, M.~Horodecki, K.~Horodecki, Quantum entanglement,
  Reviews of Modern Physics 81~(2) (2009) 865--942.
\newblock \href {https://doi.org/10.1103/RevModPhys.81.865}
  {\path{doi:10.1103/RevModPhys.81.865}}.

\bibitem{Leuenberger2001}
M.~N. Leuenberger, D.~Loss, Quantum computing in molecular magnets, Nature
  410~(6830) (2001) 789--793.
\newblock \href {https://doi.org/10.1038/35071024}
  {\path{doi:10.1038/35071024}}.

\bibitem{Gaita-Arino2019}
A.~{Gaita-Ari{\~n}o}, F.~Luis, S.~Hill, E.~Coronado, Molecular spins for
  quantum computation, Nature Chemistry 11~(4) (2019) 301--309.
\newblock \href {https://doi.org/10.1038/s41557-019-0232-y}
  {\path{doi:10.1038/s41557-019-0232-y}}.

\bibitem{Coronado2020}
E.~Coronado, Molecular magnetism: From chemical design to spin control in
  molecules, materials and devices, Nature Reviews Materials 5~(2) (2020)
  87--104.
\newblock \href {https://doi.org/10.1038/s41578-019-0146-8}
  {\path{doi:10.1038/s41578-019-0146-8}}.

\bibitem{Gatteschi1993}
D.~Gatteschi, B.~Tsukerblatt, A.~L. Barra, L.~C. Brunel, A.~Mueller,
  J.~Doering, Magnetic properties of isostructural dodecanuclear
  polyoxovanadates with six and eight vanadium({{IV}}) ions, Inorganic
  Chemistry 32~(10) (1993) 2114--2117.
\newblock \href {https://doi.org/10.1021/ic00062a039}
  {\path{doi:10.1021/ic00062a039}}.

\bibitem{Basler2002}
R.~Basler, G.~Chaboussant, A.~Sieber, H.~Andres, M.~Murrie, P.~K{\"o}gerler,
  H.~B{\"o}gge, D.~C. Crans, E.~Krickemeyer, S.~Janssen, H.~Mutka,
  A.~M{\"u}ller, H.-U. G{\"u}del, Inelastic {{Neutron Scattering}} on {{Three
  Mixed}}-{{Valence Dodecanuclear Polyoxovanadate Clusters}},
  Inorganic Chemistry 41~(22) (2002) 5675--5685.
\newblock \href {https://doi.org/10.1021/ic0202099}
  {\path{doi:10.1021/ic0202099}}.

\bibitem{Basler2002a}
R.~Basler, G.~Chaboussant, H.~Andres, P.~K{\"o}gerler, E.~Krickemeier,
  H.~B{\"o}gge, H.~Mutka, A.~M{\"u}ller, H.-U. G{\"u}del, Inelastic neutron
  scattering on a mixed-valence dodecanuclear polyoxovanadate cluster, Applied
  Physics A 74~(1) (2002) s734--s736.
\newblock \href {https://doi.org/10.1007/s003390201577}
  {\path{doi:10.1007/s003390201577}}.

\bibitem{Tribedi2006}
A.~Tribedi, I.~Bose, Entangled spin clusters: {{Some}} special features,
  Physical Review A 74~(1) (2006) 012314.
\newblock \href {https://doi.org/10.1103/PhysRevA.74.012314}
  {\path{doi:10.1103/PhysRevA.74.012314}}.

\bibitem{He2006}
M.-M. He, C.-T. Xu, J.~Q. Liang, Thermal and ground-state entanglement in the
  supermolecular dimer [{{Mn$_4$}}]$_2$, Physics Letters A 358~(5) (2006) 381--385.
\newblock \href {https://doi.org/10.1016/j.physleta.2006.05.089}
  {\path{doi:10.1016/j.physleta.2006.05.089}}.

\bibitem{Candini2010}
A.~Candini, G.~Lorusso, F.~Troiani, A.~Ghirri, S.~Carretta, P.~Santini,
  G.~Amoretti, C.~Muryn, F.~Tuna, G.~Timco, E.~J.~L. McInnes, R.~E.~P.
  Winpenny, W.~Wernsdorfer, M.~Affronte, Entanglement in {{Supramolecular Spin
  Systems}} of {{Two Weakly Coupled Antiferromagnetic Rings}}
  ({{Purple}}-Cr$_7$Ni),
  Physical Review Letters 104~(3) (2010) 037203.
\newblock \href {https://doi.org/10.1103/PhysRevLett.104.037203}
  {\path{doi:10.1103/PhysRevLett.104.037203}}.

\bibitem{Abgaryan2011}
V.~S. Abgaryan, N.~S. Ananikian, L.~N. Ananikyan, A.~N. Kocharian, Phase
  transitions and entanglement properties in spin-1 {{Heisenberg}} clusters
  with single-ion anisotropy, Physica Scripta 83~(5) (2011) 055702.
\newblock \href {https://doi.org/10.1088/0031-8949/83/05/055702}
  {\path{doi:10.1088/0031-8949/83/05/055702}}.

\bibitem{Yurishchev2011}
M.~A. Yurishchev, Quantum discord in spin-cluster materials, Physical Review B
  84~(2) (2011) 024418.
\newblock \href {https://doi.org/10.1103/PhysRevB.84.024418}
  {\path{doi:10.1103/PhysRevB.84.024418}}.

\bibitem{Reis2012}
M.~S. Reis, S.~Soriano, A.~M. dos Santos, B.~C. Sales, D.~O. {Soares-Pinto},
  P.~Brand{\~a}o, Evidence for entanglement at high temperatures in an
  engineered molecular magnet, EPL (Europhysics Letters) 100~(5) (2012) 50001.
\newblock \href {https://doi.org/10.1209/0295-5075/100/50001}
  {\path{doi:10.1209/0295-5075/100/50001}}.

\bibitem{Das2013}
D.~Das, H.~Singh, T.~Chakraborty, R.~K. Gopal, C.~Mitra, Experimental detection
  of quantum information sharing and its quantification in quantum spin
  systems, New Journal of Physics 15~(1) (2013) 013047.
\newblock \href {https://doi.org/10.1088/1367-2630/15/1/013047}
  {\path{doi:10.1088/1367-2630/15/1/013047}}.

\bibitem{Chakraborty2014}
T.~Chakraborty, H.~Singh, C.~Mitra, Signature of quantum entanglement in
  {{NH$_4$CuPO$_4$}}{$\cdot$}{{H$_2$O}}, Journal of Applied Physics 115~(3) (2014)
  034909.
\newblock \href {https://doi.org/10.1063/1.4861732}
  {\path{doi:10.1063/1.4861732}}.

\bibitem{Irons2017}
H.~R. Irons, J.~Quintanilla, T.~G. Perring, L.~Amico, G.~Aeppli, Control of
  entanglement transitions in quantum spin clusters, Physical Review B 96~(22)
  (2017) 224408.
\newblock \href {https://doi.org/10.1103/PhysRevB.96.224408}
  {\path{doi:10.1103/PhysRevB.96.224408}}.

\bibitem{Garlatti2017}
E.~Garlatti, T.~Guidi, S.~Ansbro, P.~Santini, G.~Amoretti, J.~Ollivier,
  H.~Mutka, G.~Timco, I.~J. {Vitorica-Yrezabal}, G.~F.~S. Whitehead, R.~E.~P.
  Winpenny, S.~Carretta, Portraying entanglement between molecular qubits with
  four-dimensional inelastic neutron scattering, Nature Communications 8 (2017)
  14543.
\newblock \href {https://doi.org/10.1038/ncomms14543}
  {\path{doi:10.1038/ncomms14543}}.

\bibitem{Cencarikova2020}
H.~{\v C}en{\v c}arikov{\'a}, J.~Stre{\v c}ka, Unconventional strengthening of
  the bipartite entanglement of a mixed spin-(1/2,1) {{Heisenberg}} dimer
  achieved through {{Zeeman}} splitting, Physical Review B 102~(18) (2020)
  184419.
\newblock \href {https://doi.org/10.1103/PhysRevB.102.184419}
  {\path{doi:10.1103/PhysRevB.102.184419}}.

\bibitem{Ghannadan2021}
A.~Ghannadan, J.~Stre{\v c}ka, Magnetic-{{Field}}-{{Orientation Dependent
  Thermal Entanglement}} of a {{Spin}}-1 {{Heisenberg Dimer}}: {{The Case
  Study}} of {{Dinuclear Nickel Complex}} with an {{Uniaxial Single}}-{{Ion
  Anisotropy}}, Molecules 26~(11) (2021) 3420.
\newblock \href {https://doi.org/10.3390/molecules26113420}
  {\path{doi:10.3390/molecules26113420}}.

\bibitem{Wang2001}
X.~Wang, H.~Fu, A.~I. Solomon, Thermal entanglement in three-qubit
  {{Heisenberg}} models, Journal of Physics A: Mathematical and General 34~(50)
  (2001) 11307--11320.
\newblock \href {https://doi.org/10.1088/0305-4470/34/50/312}
  {\path{doi:10.1088/0305-4470/34/50/312}}.

\bibitem{Bose2005}
I.~Bose, A.~Tribedi, Thermal entanglement properties of small spin clusters,
  Physical Review A 72~(2) (2005) 022314.
\newblock \href {https://doi.org/10.1103/PhysRevA.72.022314}
  {\path{doi:10.1103/PhysRevA.72.022314}}.

\bibitem{Pal2009}
A.~K. Pal, I.~Bose, Entanglement in a molecular three-qubit system, Journal of
  Physics: Condensed Matter 22~(1) (2009) 016004.
\newblock \href {https://doi.org/10.1088/0953-8984/22/1/016004}
  {\path{doi:10.1088/0953-8984/22/1/016004}}.

\bibitem{Pal2011}
A.~K. Pal, I.~Bose, Quantum discord in the ground and thermal states of spin
  clusters, Journal of Physics B: Atomic, Molecular and Optical Physics 44~(4)
  (2011) 045101.
\newblock \href {https://doi.org/10.1088/0953-4075/44/4/045101}
  {\path{doi:10.1088/0953-4075/44/4/045101}}.

\bibitem{Cima2016}
O.~M.~D. Cima, D.~H.~T. Franco, S.~L.~L. da~Silva, Quantum entanglement in
  trimer spin-1/2 {{Heisenberg}} chains with antiferromagnetic coupling,
  Quantum Studies: Mathematics and Foundations 3~(1) (2016) 57--63.
\newblock \href {https://doi.org/10.1007/s40509-015-0059-1}
  {\path{doi:10.1007/s40509-015-0059-1}}.

\bibitem{Anza2010}
F.~Anz{\`a}, B.~Militello, A.~Messina, Tripartite thermal correlations in an
  inhomogeneous spin\textendash star system, Journal of Physics B: Atomic,
  Molecular and Optical Physics 43~(20) (2010) 205501.
\newblock \href {https://doi.org/10.1088/0953-4075/43/20/205501}
  {\path{doi:10.1088/0953-4075/43/20/205501}}.

\bibitem{Karlova2020}
K.~Kar{\v l}ov{\'a}, J.~Stre{\v c}ka, Interplay of {{Bipartite Entanglement}}
  between {{Two Geometrically Inequivalent Spin Pairs}} of a {{Spin}}-1/2
  {{Heisenberg Distorted Tetrahedron}}, Acta Physica Polonica A 137~(5) (2020)
  595--597.
\newblock \href {https://doi.org/10.12693/APhysPolA.137.595}
  {\path{doi:10.12693/APhysPolA.137.595}}.

\bibitem{Deb2017}
M.~Deb, A.~K. Ghosh, Thermally stable multipartite entanglements in the
  frustrated {{Heisenberg}} hexagon, The European Physical Journal D 71~(6)
  (2017) 173.
\newblock \href {https://doi.org/10.1140/epjd/e2017-80179-5}
  {\path{doi:10.1140/epjd/e2017-80179-5}}.

\bibitem{Wang2002}
X.~Wang, Threshold temperature for pairwise and many-particle thermal
  entanglement in the isotropic {{Heisenberg}} model, Physical Review A 66~(4)
  (2002) 044305.
\newblock \href {https://doi.org/10.1103/PhysRevA.66.044305}
  {\path{doi:10.1103/PhysRevA.66.044305}}.

\bibitem{Kozlowski2015}
P.~Koz{\l}owski, Frustration and quantum entanglement in odd-membered
  ring-shaped chromium nanomagnets, Physical Review B 91~(17) (2015) 174432.
\newblock \href {https://doi.org/10.1103/PhysRevB.91.174432}
  {\path{doi:10.1103/PhysRevB.91.174432}}.

\bibitem{Dyszel2021}
P.~Dyszel, J.~T. Haraldsen, Thermodynamics of {{General Heisenberg Spin
  Tetramers Composed}} of {{Coupled Quantum Dimers}}, Magnetochemistry 7~(2)
  (2021) 29.
\newblock \href {https://doi.org/10.3390/magnetochemistry7020029}
  {\path{doi:10.3390/magnetochemistry7020029}}.

\bibitem{Ma2011}
X.-s. Ma, B.~Dakic, W.~Naylor, A.~Zeilinger, P.~Walther, Quantum simulation of
  the wavefunction to probe frustrated {{Heisenberg}} spin systems, Nature
  Physics 7~(5) (2011) 399--405.
\newblock \href {https://doi.org/10.1038/nphys1919}
  {\path{doi:10.1038/nphys1919}}.

\bibitem{Szalowski2020}
K.~Sza{\l}owski, Low-{{Temperature Magnetocaloric Properties}} of {{V12
  Polyoxovanadate Molecular Magnet}}: {{A Theoretical Study}}, Materials
  13~(19) (2020) 4399.
\newblock \href {https://doi.org/10.3390/ma13194399}
  {\path{doi:10.3390/ma13194399}}.

\bibitem{Procissi2004}
D.~Procissi, A.~Shastri, I.~Rousochatzakis, M.~Al~Rifai, P.~K{\"o}gerler,
  M.~Luban, B.~J. Suh, F.~Borsa, Magnetic susceptibility and spin dynamics of a
  polyoxovanadate cluster: {{A}} proton {{NMR}} study of a model spin tetramer,
  Physical Review B 69~(9) (2004) 094436.
\newblock \href {https://doi.org/10.1103/PhysRevB.69.094436}
  {\path{doi:10.1103/PhysRevB.69.094436}}.

\bibitem{Plascak2018}
J.~A. Plascak, Ensemble thermodynamic potentials of magnetic systems, Journal
  of Magnetism and Magnetic Materials 468 (2018) 224--229.
\newblock \href {https://doi.org/10.1016/j.jmmm.2018.08.014}
  {\path{doi:10.1016/j.jmmm.2018.08.014}}.

\bibitem{Hill1997}
S.~Hill, W.~K. Wootters, Entanglement of a {{Pair}} of {{Quantum Bits}},
  Physical Review Letters 78~(26) (1997) 5022--5025.
\newblock \href {https://doi.org/10.1103/PhysRevLett.78.5022}
  {\path{doi:10.1103/PhysRevLett.78.5022}}.

\bibitem{Wootters1998}
W.~K. Wootters, Entanglement of {{Formation}} of an {{Arbitrary State}} of
  {{Two Qubits}}, Physical Review Letters 80~(10) (1998) 2245--2248.
\newblock \href {https://doi.org/10.1103/PhysRevLett.80.2245}
  {\path{doi:10.1103/PhysRevLett.80.2245}}.

\bibitem{Jozsa1994}
R.~Jozsa, Fidelity for {{Mixed Quantum States}}, Journal of Modern Optics
  41~(12) (1994) 2315--2323.
\newblock \href {https://doi.org/10.1080/09500349414552171}
  {\path{doi:10.1080/09500349414552171}}.

\bibitem{Greenberger1989}
D.~M. Greenberger, M.~A. Horne, A.~Zeilinger, Going {{Beyond Bell}}'s
  {{Theorem}}, in: M.~Kafatos (Ed.), Bell's {{Theorem}}, {{Quantum Theory}} and
  {{Conceptions}} of the {{Universe}}, Fundamental {{Theories}} of {{Physics}},
  {Springer Netherlands}, {Dordrecht}, 1989, pp. 69--72.
\newblock \href {https://doi.org/10.1007/978-94-017-0849-4_10}
  {\path{doi:10.1007/978-94-017-0849-4_10}}.

\bibitem{Greenberger1990}
D.~M. Greenberger, M.~A. Horne, A.~Shimony, A.~Zeilinger, Bell's theorem
  without inequalities, American Journal of Physics 58~(12) (1990) 1131--1143.
\newblock \href {https://doi.org/10.1119/1.16243} {\path{doi:10.1119/1.16243}}.

\bibitem{Greenberger2009}
D.~M. Greenberger, {{GHZ}}
  ({{Greenberger}}\textemdash{{Horne}}\textemdash{{Zeilinger}}) {{Theorem}} and
  {{GHZ States}}, in: D.~Greenberger, K.~Hentschel, F.~Weinert (Eds.),
  Compendium of {{Quantum Physics}}, {Springer}, {Berlin, Heidelberg}, 2009,
  pp. 258--263.
\newblock \href {https://doi.org/10.1007/978-3-540-70626-7_78}
  {\path{doi:10.1007/978-3-540-70626-7_78}}.

\bibitem{Sackett2000}
C.~A. Sackett, D.~Kielpinski, B.~E. King, C.~Langer, V.~Meyer, C.~J. Myatt,
  M.~Rowe, Q.~A. Turchette, W.~M. Itano, D.~J. Wineland, C.~Monroe,
  Experimental entanglement of four particles, Nature 404~(6775) (2000)
  256--259.
\newblock \href {https://doi.org/10.1038/35005011}
  {\path{doi:10.1038/35005011}}.

\bibitem{WolframResearchInc.2010}
{Wolfram Research, Inc.}, Mathematica, version 8.0, {Champaign,
  Illinois}, 2010.

\bibitem{Bengtsson2017}
Multipartite entanglement, in: I.~Bengtsson, K.~{\.Z}yczkowski, Geometry
  of {{Quantum States}}: {{An Introduction}} to {{Quantum Entanglement}}, 2nd
  Edition, {Cambridge University Press}, {Cambridge}, 2017, pp. 493--543.
\newblock \href {https://doi.org/10.1017/9781139207010.018}
  {\path{doi:10.1017/9781139207010.018}}.

\bibitem{Siloi2012}
I.~Siloi, F.~Troiani, Towards the chemical tuning of entanglement in molecular
  nanomagnets, Physical Review B 86~(22) (2012) 224404.
\newblock \href {https://doi.org/10.1103/PhysRevB.86.224404}
  {\path{doi:10.1103/PhysRevB.86.224404}}.

\bibitem{Ghirri2015}
A.~Ghirri, F.~Troiani, M.~Affronte, Quantum {{Computation}} with {{Molecular
  Nanomagnets}}: {{Achievements}}, {{Challenges}}, and {{New Trends}}, in:
  S.~Gao (Ed.), Molecular {{Nanomagnets}} and {{Related Phenomena}}, Structure
  and {{Bonding}}, {Springer}, {Berlin, Heidelberg}, 2015, pp. 383--430.
\newblock \href {https://doi.org/10.1007/430_2014_145}
  {\path{doi:10.1007/430_2014_145}}.

\bibitem{Lehmann2007}
J.~Lehmann, A.~{Gaita-Ari{\=n}o}, E.~Coronado, D.~Loss, Spin qubits with
  electrically gated polyoxometalate molecules, Nature Nanotechnology 2~(5)
  (2007) 312--317.
\newblock \href {https://doi.org/10.1038/nnano.2007.110}
  {\path{doi:10.1038/nnano.2007.110}}.

\bibitem{Clemente-Juan2012}
J.~M. {Clemente-Juan}, E.~Coronado, A.~{Gaita-Ari{\~n}o}, Magnetic
  polyoxometalates: From molecular magnetism to molecular spintronics and
  quantum computing, Chemical Society Reviews 41~(22) (2012) 7464--7478.
\newblock \href {https://doi.org/10.1039/C2CS35205B}
  {\path{doi:10.1039/C2CS35205B}}.

\bibitem{Baldovi2017}
J.~J. Baldov{\'i}, S.~{Cardona-Serra}, A.~{Gaita-Ari{\~n}o}, E.~Coronado,
  Chapter {{Eight}} - {{Design}} of {{Magnetic Polyoxometalates}} for
  {{Molecular Spintronics}} and as {{Spin Qubits}}, in: R.~{van Eldik},
  L.~Cronin (Eds.), Advances in {{Inorganic Chemistry}}, Vol.~69 of
  Polyoxometalate {{Chemistry}}, {Academic Press}, 2017, pp. 213--249.
\newblock \href {https://doi.org/10.1016/bs.adioch.2016.12.003}
  {\path{doi:10.1016/bs.adioch.2016.12.003}}.

\bibitem{Scheie2021}
A.~Scheie, P.~Laurell, A.~M. Samarakoon, B.~Lake, S.~E. Nagler, G.~E. Granroth,
  S.~Okamoto, G.~Alvarez, D.~A. Tennant, Witnessing entanglement in quantum
  magnets using neutron scattering, Physical Review B 103~(22) (2021) 224434.
\newblock \href {https://doi.org/10.1103/PhysRevB.103.224434}
  {\path{doi:10.1103/PhysRevB.103.224434}}.

\bibitem{Laurell2021}
P.~Laurell, A.~Scheie, C.~J. Mukherjee, M.~M. Koza, M.~Enderle, Z.~Tylczynski,
  S.~Okamoto, R.~Coldea, D.~A. Tennant, G.~Alvarez, Quantifying and
  {{Controlling Entanglement}} in the {{Quantum Magnet}} Cs$_2$CoCl$_4$, Physical
  Review Letters 127~(3) (2021) 037201.
\newblock \href {https://doi.org/10.1103/PhysRevLett.127.037201}
  {\path{doi:10.1103/PhysRevLett.127.037201}}.

\bibitem{Belik2007}
A.~A. Belik, H.-J. Koo, M.-H. Whangbo, N.~Tsujii, P.~Naumov,
  E.~{Takayama-Muromachi}, Magnetic {{Properties}} of {{Synthetic Libethenite
  Cu$_2$PO$_4$OH}}:\, a {{New Spin}}-{{Gap System}}, Inorganic Chemistry 46~(21)
  (2007) 8684--8689.
\newblock \href {https://doi.org/10.1021/ic7008418}
  {\path{doi:10.1021/ic7008418}}.

\end{thebibliography}

\appendix
\section{\label{a}The ground states and critical magnetic fields for approximate tetramer models}

\subsection{\label{a1}Ground state with $S=0$, isotropic tetramer}

For fully isotropic tetramer, the ground state with $S=0$ takes the form of \cite{Wang2002}:
\begin{equation}
\left|\psi_{S=0}\right\rangle=\frac{1}{2\sqrt{3}}\left(\left|\uparrow\uparrow\downarrow\downarrow\right\rangle+\left|\uparrow\downarrow\downarrow\uparrow\right\rangle+\left|\downarrow\uparrow\uparrow\downarrow\right\rangle+\left|\downarrow\downarrow\uparrow\uparrow\right\rangle-2\left|\uparrow\downarrow\uparrow\downarrow\right\rangle-2\left|\downarrow\uparrow\downarrow\uparrow\right\rangle\right),
\end{equation}
with the energy of
\begin{equation}
E_{S=0}=2J.
\end{equation}

\subsection{\label{a2}Ground state with $S=0$, tetramer with real-space anisotropy}

For such a model, the ground state with spin $S=0$ is:
\begin{align}
\left|\psi_{S=0}\right\rangle=\frac{1}{c}
   &\left[\left|\uparrow\uparrow\downarrow\downarrow\right\rangle+\left|\downarrow\downarrow\uparrow\uparrow\right\rangle + \frac{1}{\sqrt{1-x+x^2}-x} \left( \left|\uparrow\downarrow\downarrow\uparrow\right\rangle+  \left|\downarrow\uparrow\uparrow\downarrow\right\rangle\right)\right.\nonumber\\   &\left.   
-\frac{1+\sqrt{1-x+x^2}}{1-x} \left(\left|\uparrow\downarrow\uparrow\downarrow\right\rangle  +\left|\downarrow\uparrow\downarrow\uparrow\right\rangle\right)\right],
\end{align}
where $x=1-\frac{J_2}{J_1}$ and the normalization constant is:
\begin{equation}
c=2\sqrt{2\left(1-x+x^2\right)+\left(1+x\right)\sqrt{1-x+x^2}}.
\end{equation}

\subsection{\label{a3}Ground state with $S=0$, tetramer with spin-space anisotropy}
In the case of sole spin-space anisotropy of couplings in the studied tetramer, the ground state with spin $S=0$ is:
\begin{equation}
\left|\psi_{S=0}\right\rangle=\frac{1}{c}\left[\left|\uparrow\uparrow\downarrow\downarrow\right\rangle+\left|\uparrow\downarrow\downarrow\uparrow\right\rangle+\left|\downarrow\uparrow\uparrow\downarrow\right\rangle+\left|\downarrow\downarrow\uparrow\uparrow\right\rangle-\frac{1+\sqrt{1+8\left(\frac{J^{\perp}}{J^{z}}\right)^2}}{2\left(\frac{J^{\perp}}{J^{z}}\right)}\left(\left|\uparrow\downarrow\uparrow\downarrow\right\rangle+\left|\downarrow\uparrow\downarrow\uparrow\right\rangle\right)\right],
\end{equation}
where the normalization constant is:
\begin{equation}
c=\sqrt{8+\frac{1+\sqrt{1+8\left(\frac{J^{\perp}}{J^{z}}\right)^2}}{\left(\frac{J^{\perp}}{J^{z}}\right)^2}}.
\end{equation}

\subsection{\label{a4}Ground state with $S=1$, all the models for tetramer}

For all the theoretical models considered in the section \ref{theory}, the state takes the following form \cite{Wang2002}:
\begin{equation}
\left|\psi_{S=1}\right\rangle=\frac{1}{2}\left(\left|\downarrow\uparrow\uparrow\uparrow\right\rangle-\left|\uparrow\downarrow\uparrow\uparrow\right\rangle+\left|\uparrow\uparrow\downarrow\uparrow\right\rangle-\left|\uparrow\uparrow\uparrow\downarrow\right\rangle\right)
\end{equation}

\subsection{\label{a5}Ground state with $S=2$, all the models for tetramer}
The state takes the following form \cite{Wang2002} for all the interaction models considered:
\begin{equation}
\left|\psi_{S=2}\right\rangle=\left|\uparrow\uparrow\uparrow\uparrow\right\rangle.
\end{equation}

\subsection{\label{a6}Critical magnetic fields, isotropic tetramer}
The critical field for transition between the ground state with $S=0$ and with $S=1$ is equal to
\begin{equation}
B_{c,1}=\left|J\right|,
\end{equation}
while the critical field for transition between the ground state with $S=1$ and with $S=2$ is equal to
\begin{equation}
B_{c,2}=2\left|J\right|.
\end{equation}

\subsection{\label{a7}Critical magnetic fields, tetramer with real-space anisotropy}
The critical field for transition between the ground state with $S=0$ and with $S=1$ is equal to
\begin{equation}
B_{c,1}=\left|J_{1}\right| \sqrt{\frac{J_2}{J_1}+ \left(1-\frac{J_2}{J_1}\right)^2},
\end{equation}
with the series expansion in the following form:
\begin{equation}
B_{c,1}\simeq \left|J_{1}\right|\left[1-\frac{1}{2}\left(1-\frac{J_2}{J_1}\right)+\frac{3}{8}\left(1-\frac{J_2}{J_1}\right)^2+\frac{3}{16}\left(1-\frac{J_2}{J_1}\right)^3+\frac{3}{128}\left(1-\frac{J_2}{J_1}\right)^4\right].
\end{equation}
The critical field for transition between the ground state with $S=1$ and with $S=2$ is equal to
\begin{equation}
B_{c,2}=\left|J_{1}\right|\left(1+\frac{J_{2}}{J_{1}}\right).
\end{equation}

\subsection{\label{a8}Critical magnetic fields, tetramer with spin-space anisotropy}
The critical field for transition between the ground state with $S=0$ and with $S=1$ is equal to
\begin{equation}
B_{c,1}=\frac{1}{2}\left|J^{z}\right| \left(1-2\frac{J^{\perp}}{J^{z}}+\sqrt{8 \left(\frac{J^{\perp}}{J^{z}}\right)^2+1}\right),
\end{equation}
with the series expansion in the following form:
\begin{equation}
B_{c,1}\simeq \left|J^{z}\right|\left[1-\frac{1}{3}\left(1-\frac{J^{\perp}}{J^{z}}\right)+\frac{2}{27}\left(1-\frac{J^{\perp}}{J^{z}}\right)^2+\frac{16}{243}\left(1-\frac{J^{\perp}}{J^{z}}\right)^3+\frac{124}{2187}\left(1-\frac{J^{\perp}}{J^{z}}\right)^4\right].
\end{equation}
The critical field for transition between the ground state with $S=1$ and with $S=2$ is equal to
\begin{equation}
B_{c,2}=\left|J^{z}\right|\left(1+\frac{J^{\perp}}{J^{z}}\right).
\end{equation}

In V12, the values of the critical fields amount to $B_{c,1}=$13.3 T and $B_{c,2}=$ 26.1 T, according to the tetramer model including both anisotropies, parametrized by the exchange integrals taken from Ref.~\cite{Basler2002,Basler2002a}.

\end{document}